# Uncertainty promotes neuroreductionism: A behavioral online study on folk psychological causal inference from neuroimaging data

Jona Carmon[abc*], Moritz Bammel[a*], Peter Brugger[de], Bigna Lenggenhager[f]

[a] Berlin School of Mind and Brain, Humboldt-University Berlin, Berlin, Germany

[b] Humanities and Educational Science, Technical University Berlin, Berlin, Germany

[c] College of Architecture, Media and Design, University of the Arts, Berlin, Germany

[d] Rehabilitation Center Valens, Valens, Switzerland

[e] University Hospital of Psychiatry (PUK), Zurich, Switzerland

[f] Department of Psychology, University of Zurich, Zurich, Switzerland

[*] J.C. and M.B. contributed equally

Short Title: Uncertainty promotes neuroreductionism

Corresponding Author:

Prof. Dr. Bigna Lenggenhager

University of Zurich

Institute of Psychology

Cognitive Neuropsychology

Binzmühlestrasse 14, Box 9

CH-8050 Zurich

Phone: +41 44 635 72 18

Email: bigna.lenggenhager@psychologie.uzh.ch

# Abstract

**Introduction**. Increased efforts in neuroscience try to understand mental disorders as brain disorders. In the present study we investigate how common a neuroreductionist inclination is among highly educated people. In particular, we shed light on implicit presuppositions of mental disorders little is known about in the public, exemplified here by the case of Body Integrity Dysphoria (BID) that is considered a mental disorder for the first time in ICD-11.

**Methods.** Identically graphed, simulated data of mind-brain correlations were shown in three contexts with presumably different presumptions about causality. 738 highly-educated laymen rated plausibility of causality attribution from brain to mind and from mind to brain for correlations between brain structural properties and mental phenomena. We contrasted participants' plausibility ratings of causality in the contexts of commonly perceived brain-lesion induced behavior (aphasia), behavior-induced training effects (piano playing), and a newly described mental disorder (BID).

**Results.** The findings reveal the expected context-dependent modulation of causality attributions in the contexts of aphasia and piano playing. Furthermore, we observed a significant tendency to more readily attribute causal inference from brain to mind than vice versa with respect to BID.

**Conclusion.** In some contexts, exemplified here by aphasia and piano playing, unidirectional causality attributions may be justified. However, with respect to BID, we critically discuss presumably unjustified neuroreductionist inclinations under causal uncertainty. Finally, we emphasize the need for a presupposition-free approach in psychiatry.

**Introduction**

The relation between mind and brain has a long and convoluted history, both in the philosophy of mind and in neuroscience. Those who think they have overcome mind-brain dualism are often lost in positions betwixt and between, and their reasoning continues to operate dualistically, if only in covert and unacknowledged ways [1]. An overt solution of dualistic thinking is arguably offered by neuroreductionism which is here understood as reducing mental phenomena to merely phenomena of the brain. More than a century ago, Griesinger famously stated that all mental diseases are brain diseases [2], a view that has recently been echoed authoritatively [3]. Although the claim that neural structure would cause mental disorder is not made explicit, it certainly lurks behind such an assumption. Many scholars have criticized such accounts by stressing the importance of integrating somoatopsychic and socio-psychosomatic factors in a fashion which is neither brainless nor mindless [4,5].

Such debates are not only of academic interest, but the conceptualization and the assumed cause of mental disorders determine their social acceptance, classification and basic medical treatment strategies. Classification systems like the International Classification of Diseases (ICD) or the Diagnostic and Statistical Manual of Mental Disorders (DSM) are constantly updated and reflect current societal perspective on mental health as much as they influence it [6].

Here, we focus on a condition newly introduced as a mental disorder by ICD-11 (https://icd.who.int/browse11/l-m/en#/http://id.who.int/icd/entity/256572629). Labeled Body Integrity Dysphoria (BID), it designates the pathological suffering from having a nondisabled body. Its most frequent variant is specific and consists in the desire for amputation of a healthy limb. The term Xenomelia was proposed by those considering it a focal disease of the brain [7]. Others conceptualize BID as a pure sickness of the mind, i.e. an "Internet-based madness" [8].

How neuroscientific results are generally communicated to the public has been addressed by previous research and the potential impact has been discussed critically [9, 10]. The present study concerns the causal role attributed to the brain. Unjustified causal inferences from correlation are a common practice, even in highly educated individuals. In order to investigate neuroreductionist tendencies in educated, but non-expert quarters, and how much it is influenced by context, we collected plausibility ratings of suggested mind-brain and brain-mind causalities for identically depicted typical correlative neuroimaging data. We contrasted three different contexts with presumably different degrees of general knowledge about hypothesized brain-behavior relations. Like our simulated data, neuroimaging studies generally describe correlations between a neural measure (e.g. density of gray matter, blood oxygenation etc.)

and a specific behavior or a subjective state. Such correlations do not allow any causal inference [11]. Therefore, causal judgments of correlative data elucidate implicit presuppositions, either from brain to mind or vice versa. These presuppositions might depend on folk psychological intuitions such as the assumed temporal relation in the specific contexts. Especially if little additional information is provided, as in the case of a newly described disorder like BID, such causal inference plausibility judgments may inform us about participants' beliefs about brain mind relation in mental disorders.

In order to capture prevalent implicit folk psychological intuitions on the etiology of mental disorders, we deliberately refer to dualistic mental vocabulary, without intending to advocate in favor of a dualist ontological commitment with respect to the mind-body problem in the philosophy of mind. Considering mental states as potential causes does not necessarily entail a dualist conception of mind and body, since monists may regard mental states as causes as well [12]. However, we do not aim to make any contribution to philosophical debates on the ontology of mind and body. The present investigation concerns implicit causal biases with respect to brain-behavior correlations among highly educated lay people and we aim to critically examine how those presuppositions affect conceptualizing yet unknown mental disorders.

## Materials and Methods

*Study design and sample*

In the present study, we depicted the same simulated correlation in different contexts, which were induced by the labels of the axes. Aphasia after stroke, professional piano playing and xenomelia, were chosen as three representative contexts in order to evoke different assumptions in lay people regarding the temporal aspects of the event and thus bias causality judgments: brain damage leading to change in behavior (context stroke / receptive aphasia), intense musical training leading to brain alterations (context professional piano playing), and a new mental disorder, where no clear information about the temporal course is currently available (context xenomelia).

Xenomelia seems to be a particularly suited condition in order to shed light on implicit causal biases in highly educated lay people with respect to interpreting brain-behavior correlations. We deliberately did not select publicly better known diseases such as depression or schizophrenia, as it can be assumed that participants are influenced by prior knowledge about the pathogenesis of these diseases. Participants' causality attributions on a yet unknown condition, however, may inform us about implicit biases that affect how highly educated individuals will most likely conceptualize the plausibility of potential causal mechanisms of a yet poorly understood phenomenon *a priori*.

Given the clear time course of events in the first context - the cerebral infarction precedes symptoms of receptive aphasia – we expected the additional information, contained in folk-neuropsychological knowledge [13], to bias participants' plausibility judgments in favor of a causality from brain to mind (compared to that in the opposite direction). With respect to musical training, in line with both the common sense that the behavior (early and extensive piano practice) precedes potential brain alterations and scientific findings showing training-induced alterations in gray matter density [14], we expected higher mind to brain plausibility ratings. We compare these two contexts to a correlation between symptoms of xenomelia and structural properties of a circumscribed cortical surface area, loosely inspired by the first publication on neurostructural data in xenomelia [15]. We assumed no additional information justifying unidirectional judgment with respect to this largely unknown disorder, but expected higher plausibility ratings from brain to mind [16].

We recruited three independent samples of participants with third level education only, one for each context (n=253 for xenomelia, n=258 for stroke / receptive aphasia, n= 251 for piano players per sample) via the online recruiting service Prolific (https://www.prolific.co). Participants confirmed that they either have obtained a university degree (Bachelor's, Master's, or Doctoral degree), or that they were undergoing higher educational training at university. Eligibility was restricted to third level education only, in order to make sure participants are familiar with the distinction of correlation and causality, which is usually addressed in introductory classes of all university degrees.

The surveys were created using the software PsyToolkit [17,18]. In each context, participants saw the same scatter plot of a specific behavioral or subjective measure and its neural correlate, as well as a brain plot with color-coded temporoparietal junction (TPJ) (generated with BrainNet Viewer [19] and a corresponding TPJ mask (https://identifiers.org/neurovault.image:12004). The depicted data (shown in Fig. 1) thus showed scientifically plausible, but not actually collected data. The TPJ was chosen as a plausible representative neural correlate for all three contexts [20 - 22]. The only difference between the contexts was the labeling of behavioral and neural measures to the x or y axis. Furthermore, to avoid confounding effects due to the general accord that predictors are typically associated with the x-axes, the two axes were randomized within each group. Below the plot we provided a short description of each context. Plausibility ratings were assessed on a visual analog scale (VAS), and we refer to *brain primacy*, as the plausibility of the vector of causality pointing from brain to mind and to *mind primacy*, as the plausibility of the vector pointing from mind to brain. The intention of including two questions in our questionnaire (one for mind and one for brain primacy) is to allow participants to conceive the directionality of assumed brain-behavior causality as not mutually exclusive. Both the order of presenting the plausibility ratings as well as the order of the individual components of the context descriptions were randomized. Figure 1 illustrates the survey data and main questions for all three contexts.

In addition to investigating brain and mind primacy we also investigated the overall importance of the research concerning brain-mind relationships in each particular context. The overall important rating was assessed in each context by a survey question with a visual analog scale (VAS) from 0 to 100, similar to the questions on mind and brain primacy. The question on the importance rating appeared in the survey after the questions on mind and brain primacies.

Furthermore, the surveys were complemented by demographic questions aiming at collecting data for exploratory analysis. Participants were asked to indicate their cultural and religious affiliations after they had provided plausibility ratings on brain- and mind primacy. We did not formulate any specific hypothesis, but we planned to conduct an exploratory analysis on potential associations between particular causal attributions and different cultural and/or religious affiliations, if the sample had turned out to be sufficiently heterogeneous. Approximately 77% of all participants considered themselves as secular and approximately 60% selected 'Western' to describe their cultural affiliation. Since the sample sizes of the respective sub-samples of specific non-Western cultural affiliations ('Arabic', 'Indosphere', 'Sinosphere', 'Africa'), or neither secular, nor Christian religious affiliations ('Islam', 'Hinduism', 'Judaism', 'Buddhism') turned out to be too small to yield any statistically meaningful results (n<10), we did not conduct any further exploratory analysis.

Prior to data collection we had pre-registered our hypothesis and statistical analysis procedure (https://osf.io/35jpw).

The research was conducted in accordance with the World Medical Association Declaration of Helsinki. The ethics committee of the Faculty of Art and Social Sciences at the University of Zurich approved this study (Approval number 19.10.6) and all participants provided informed consent by clicking on a box to confirm that they have read and understood the study information and agree to participate.

*Statistical analysis*

Brain and mind primacy were compared by separate Wilcoxon signed rank tests for each context. Inter-context differences of mind and brain primacy were analyzed using a rank-transformed one-way ANOVA. Participants who indicated to suffer from xenomelia (n=4), had experienced receptive aphasia (n=1) as well as professional piano players (n=4) and participants who did not provide any causality judgements were excluded from the analysis. The final samples comprised 244 (159 females; age 22.6 +/- 3.8) participants for the context of xenomelia, 249 (131 females; age 23.0 +/- 4.7) participants for the context of receptive aphasia and 245 (139 females; age 23.4 +/- 4.9) participants for the context of piano players.

**Results**

In accordance with our prediction, we found a significant brain primacy effect in the context stroke / receptive aphasia ($p < 0.0001$, effect size $r = 0.79$) and a significant mind primacy effect in the context professional piano playing ($p < 0.0001$, effect size $r = -0.36$). Furthermore, as expected, the causal interpretation from brain to mind was significantly more prominent than from mind to brain in the context xenomelia ($p < 0.0001$, effect size $r = 0.58$). Even though we argued that the third context lacks additional information that would justify unidirectional causal judgment, our data indicate a significant brain primacy in the context xenomelia (shown in Fig. 2). The inter-context differences of brain and mind primacy were also reflected in the significant results of the ANOVA on ranks (inter-context brain primacy $p < 0.0001$, inter-context mind primacy $p < 0.0001$; shown in Fig. 2).

An exploratory correlative analysis further revealed a statistically significant negative correlation between mind and brain primacy in the group stroke / aphasia ($r = -0.22$, $p < 0.001$). However, subsequent visual analysis did not reveal any meaningful interpretation of this correlation. The other two contexts did not show any significant correlation between mind and brain primacy (BID $r = -0.07$, $p = 0.25$; piano players $r = -0.04$, $p = 0.53$).

The overall importance ratings were also evaluated using an ANOVA on ranks and the results show significant differences between the contexts ($p << 0.0001$). The importance rating is lowest for the context professional piano players (median = 71) and highest for the context stroke / receptive aphasia (median = 88). The results of the importance rating of the context of xenomelia are in between (median = 78).

**Discussion**

The results suggest that the sample of highly educated lay people here investigated indeed adapt their (unjustified) causality judgements to the particular context they are confronted with. This indicates that our measure of primacy effects well captures the presuppositions of the participants. In the case of Body Integrity Dysphoria, a still largely unknown clinical condition, the data reveal a clear preference for a brain-to-mind directed causality, indicating that participants assumed the phenomenal and behavioral properties to be a consequence, rather than a cause of brain alterations. A similar finding was previously reported by Brugger and colleagues [16]; however, the present investigations highlight that implicit assumptions about a correlation that importantly bias the misattribution to an underlying cause significantly depend on the context. Critically, we draw attention to context dependent differences in *justifying* those causal misattributions.

We deliberately chose BID as a presumably unknown psychiatric disorder in order to gain insight into how individuals attribute causality to the brain under uncertainty, thus revealing an implicit bias. In this sense, we consider BID to be representative of publicly presumably unknown psychiatric conditions, but one must not generalize the present findings on BID to publicly well known psychiatric diseases such as depression or schizophrenia. Further research could address causality attributions with respect to known psychiatric disorders. However, studying publicly well known psychiatric conditions would primarily capture individuals' prior scientific knowledge, rather than their implicit attitudes towards those diseases.

The present work shows that educated lay people do infer from correlation to causality in a way that suggests neuroreductionism, popular among experts of past and contemporary neuropsychiatry [2,3]. We argue that these beliefs should be carefully discussed and critically reflected upon. Consider the pathogenesis of psychiatric diseases. Mental disorders have been identified by phenomenological reports of mental suffering in the first place [2]. Only afterwards, neural mechanisms which correlate with mental discomfort may be identified [2], and might be potentially developed into biomarkers. The neuroreductionist stance tries to replace the definition of mental disorders by precisely those neural correlates that could not have been identified without phenomenological reports of people's suffering. Additionally, cultural variations play an important role in defining abnormality, thus, sociocultural factors are another variable beyond the current scope of neurological biomarkers that are of paramount importance in order to understand and define mental disorders [5].

Although we interpret the present findings as being indicative of neuroreductinist tendencies under uncertainty among highly educated lay people, we did not inquire participants' detailed causal reasoning. In consequence, the present data do not allow making any inference about participants' conception of what is known as the mind-body problem in the philosophy of mind. Based on the exploratory analysis of potential correlations between mind and brain primacy, we conclude that the absence of a consistent meaningful negative correlation of mind and brain primacy across contexts seems to indicate that participants predominantly do not share a mutually exclusive unidirectional view of brain-behavior causality. Even though we found a statistically significant negative correlation in the context stroke / aphasia ($r = -0.22$), subsequent visual analysis revealed that it does not justify the inference that participants conceived mind and brain primacy as mutually exclusive in that context.

Furthermore, significant alterations in overall importance ratings are similar to differences in brain primacy across the different contexts. We speculate, however, that differences in overall importance ratings most likely reflect the clinical relevance of research in that particular context. Research on piano playing appears to be clinically less relevant than research on stroke / receptive aphasia.

How neuroscience is perceived by the public covers a wide range of issues and our study is limited to only one relevant aspect. One may ask what other factors might have affected participants' causality attributions, besides individuals' prior knowledge on temporal sequences of events and their implicit conceptions on the etiology of yet poorly characterized psychiatric diseases. For example, there is some literature on how brain plots nudge the scientific evaluation of lay people [23 - 27], but in the present study we did not further investigate how different plots may alter brain and mind primacy. The brain plot we presented in addition to the simulated data was consistent across contexts and it was solely presented in order to plausibly mimic and ensure consistency with conventional neuroimaging studies.

With respect to BID, the current neuroreductionist stance might hinder a full understanding of mental disorders as it fails to take psychosocial and phenomenological components into account. A proper social neuroscience approach seems to be more suitable in order to understand the mechanisms associated with symptoms of BID. Measurable neural correlates of the body in the brain, perceptions of bodily features and sexual identity, as well as culture-bound norms of body appearance and modification should be integrated into a "social neuroscience view of xenomelia" [28]. Such integration would welcome the label "body integrity dysphoria" once xenomelia will finally turn into an officially acknowledged mental disorder. After all, it emphasizes an individual's suffering, free of presuppositions about any primacy of brain or mind. This emphasis is precisely what psychiatry's role is about.

**Acknowledgement**

We thank the Swiss Study Foundation for support and encouragement, Maria Lampe for contributing to the study idea and Jasmine Ho for proofreading.

**Statement of Ethics**

The authors assert that all procedures contributing to this work comply with the ethical standards of the relevant national and institutional committees on human experimentation and with the Helsinki Declaration of 1975, as revised in 2008.

The ethics committee of the Faculty of Art and Social Sciences at the University of Zurich approved this study (Approval number 19.10.6) and all participants provided informed consent.

**Conflict of Interest Statement**

The authors have no conflicts of interest to declare.

**Funding Sources**

This research received no specific grant from any funding agency, commercial or not-for-profit sectors.

**Data Availability Statement**

The data that support the findings of this study are openly available on the Open Science Framework at https://doi.org/10.17605/OSF.IO/FCZ72, reference number FCZ72.

**Author Contributions**

B.L., P.B. and J.C. conceived the idea. J.C and M.B. created the survey, collected the data, analyzed the data, designed the figures and wrote the first manuscript. B.L. and P.B. supervised each working step, guided the work and revised the manuscript. All authors approved the final version of the manuscript.

# Figures

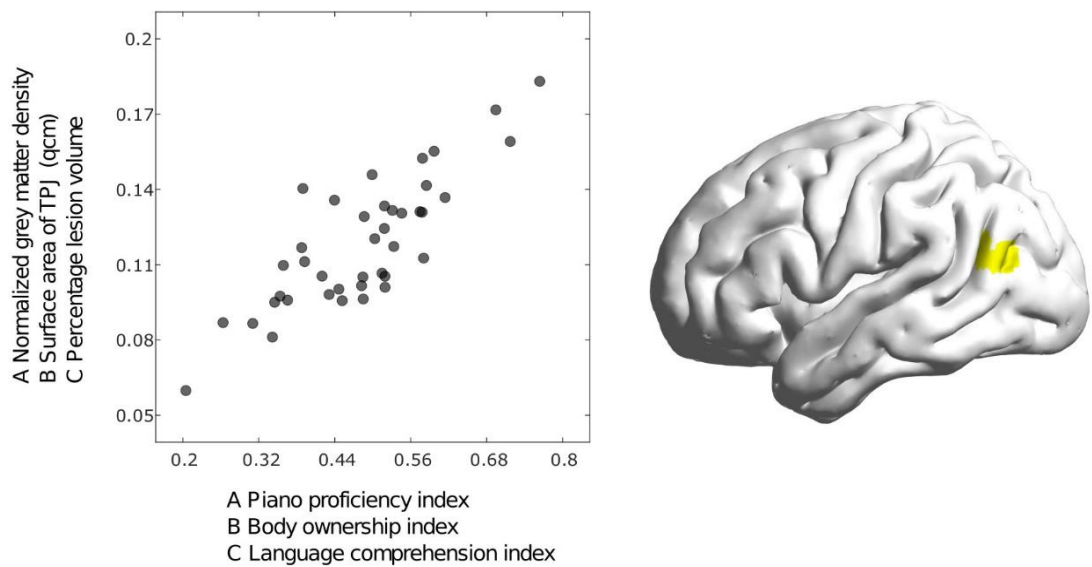

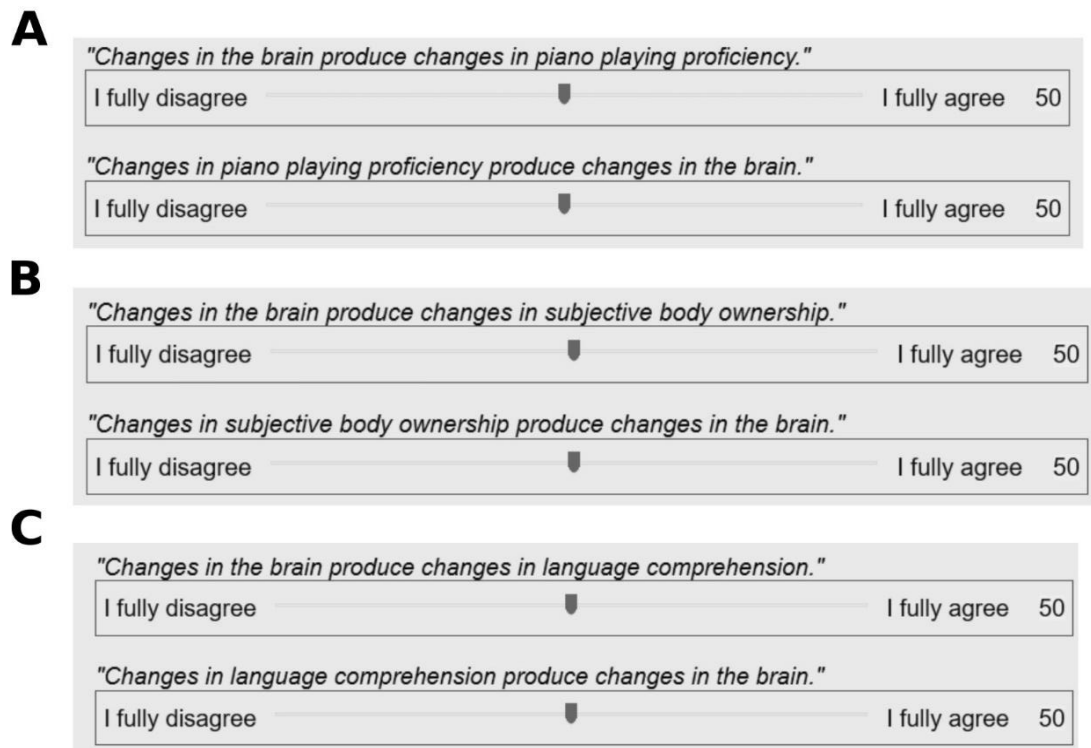

Fig. 1. **Survey Plots**. The same scatter and brain plots are displayed for all three contexts. **(A)** shows the labeling and survey questions for the context professional piano players, **(B)** for the context xenomelia and **(C)** for the context stroke / receptive aphasia.

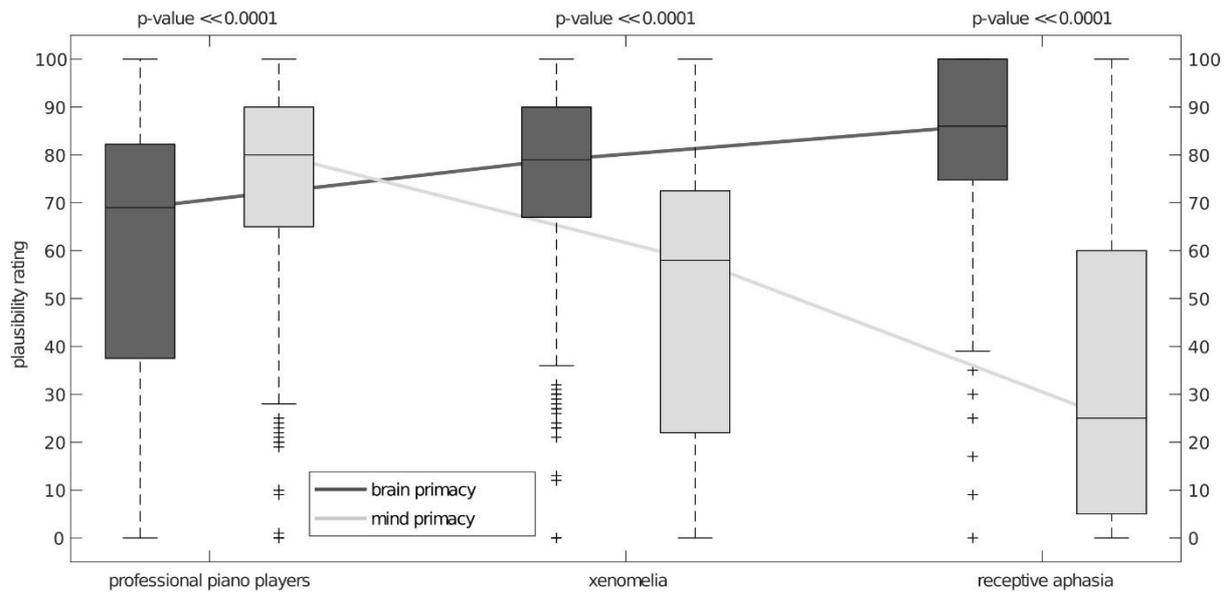

Fig. 2. **Survey Results.** The box plots show the comparison of brain and mind primacy in all three contexts. The lines show the comparison of brain and mind primacy between the contexts (ANOVA p-value of brain and mind primacy respectively << 0.0001).